\input{aipcheck}

\documentclass[
    ,final            
  ]
  {aipproc}

\layoutstyle{6x9}
\usepackage{graphicx}

\begin{document}

\title{Confinement of color: open problems and perspectives.}


\author{Adriano Di Giacomo}{
  address={Dipartimento di Fisica  Univ. PISA and INFN , 3 Largo B. Pontecorvo,56127 Pisa (ITALY)}}

\begin{abstract}
 Some basic features of  confinement are reviewed, in particular the symmetry patterns of the dual dynamics.  Open problems and possible directions of progress are discussed.
\end{abstract}

\maketitle


\section{Experiment.}

  The experimental search for quarks  was started in the 60's , immediately after the conjecture of their existence as fundamental building blocks of hadrons\cite{Gellmann}. The experimental signature is clear : they carry fractional charges  $q=\pm1/3,\pm 2/3$.
  
  No quark has ever been observed, so that experiments only provide upper limits, which are summarized in Particle Data Group reports \cite{PDG}.
  
  For the ratio of the abundance of quarks $n_q$ to the abundance of protons   $n_p$  in  nature the limit is 
  \begin{equation}
  \{ {n_q\over n_p}\}_{observed} \leq 10^{-27}
  \end{equation}
  to be compared to the expectation  $\{{n_q\over n_p}\}_{expected}\simeq 10^{-12}$ in the Standard Cosmological Model \cite{Okun}.
  
  The reduction factor is  $\simeq 10^{-15}$ !
  
  Similarly for the inclusive cross -section  $\sigma_p$  to produce a quark  or an antiquark in hadron collisions one has \cite{PDG}
  \begin{equation}
 \{\sigma_q\}_{observed}\equiv \{\sigma (p+p\to q (\bar q) + X)\}_{observed} \leq 10^{-40} cm^2
 \end{equation}
 to be compared to the expected value in perturbation theory which is of the order of the total cross section  $\sigma_T\simeq 10^{-25} cm^2$ . Again a reduction factor  $\simeq 10^{-15}$ !
 
 The  "natural " explanation is that  $n_q$ and $\sigma _q$ are strictly zero , or that confinement is an absolute property of vacuum based on symmetry.  This is similar to what happens in ordinary superconductivity: very  small experimental limits exist for the resistivity , which in fact proves to be strictly zero due to the Higgs-breaking of the $U(1)$ symmetry corresponding to the conservation of the electric charge. 
 
 The above statement has far reaching consequences: the deconfining transition can not  be a crossover if $\sigma_q$ and $n_q$ are strictly zero, since by analyticity they would be zero everywhere in the phase diagram. 
 
 An "unnatural" alternative also exists compatible with a crossover: $n_q$ and $\sigma_q$ very small but not identically zero in the confined phase and there is no order parameter. Of course the choice is not a matter of opinions , but has been made by Nature.  What we can do is to investigate the question e.g. by numerical simulations on the lattice.
\section{The deconfining phase transition}
The possible existence of a deconfining phase transition at a temperature  $T_c$ of the order of the Hagedorn temperature was first discussed in Ref.\cite{CABIBBOPARISI} . Such a transition has been observed in numerical simulations of QCD on the Lattice. 

The partition function  of $QCD$ at temperature $T$ can be written as a Feynman integral on a space- time compactified in time from $0 $  to ${1\over T}$ with periodic boundary conditions for the gluon fields, antiperiodic for the quark fields. If $L_s$ is the spatial extension of the lattice , $L_t$ the extension in time , with  $L_s\gg L_t$ ,
\begin{equation}
                            T =  {1\over {a(\beta,m)L_t}}
 \end{equation}
 where $a(\beta,m)$ is the lattice spacing in physical units, which depends on the coupling constant $g^2$  ($\beta={2N\over g^2}$)  and on the quark masses $m$ , which are the inputs of the simulations.
                          
The real problem, both on the lattice and in experiment, is to give an operational definition of confined and deconfined.

In the quenched theory (no dynamical quarks) the Polyakov criterion is usually adopted. The static potential between a $q \bar q$ pair , $V(r)$ is related to the correlator V(r) of two Polyakov lines by the equation
\begin{equation}
V(r)  =  - T lnD(r)
\end{equation}
  By cluster property
  \begin{equation}
  D(r) = < L^{\dagger}(\vec r) L(0)> \approx_{r\to \infty}|<L>|^2  + k . exp(-{\sigma r \over T})
  \end{equation}
 $ L(\vec r) $ is the Polyakov line , or the parallel transport along the time axis across the lattice.
 
 If  $<L> = 0$           $V(r) \approx_{r\to \infty} \sigma r  $                        (Confinement).
 
 If  $<L> \neq 0$      $V(r) \approx_{r\to \infty} constant  $                      (Deconfinement).
 
 $<L>$  is an order parameter, the symmetry being  $Z_3$.   A phase transition is found  on the lattice at a temperature $T_c\simeq 270 Mev$  from a phase   $T\leq T_c$ where  $<L> = 0$ to a phase at $T>0$
 where  $<L>\neq 0$ . Finite size scaling analysis shows that the transition is weak first order.
 
   In the presence of dynamical quarks  the symmetry $Z_3$ is explicitly broken  and the string breaks
   at large enough distances converting the energy into the formation of quark pairs. No clear criterion exists to define confined and deconfined. 
   
   Fig.1 shows the phase diagram for $N_f=2$ and $u$ and $d$
   quarks with the same  mass $m$.
   \begin{figure}[!h]
  \includegraphics[height=.3\textheight]{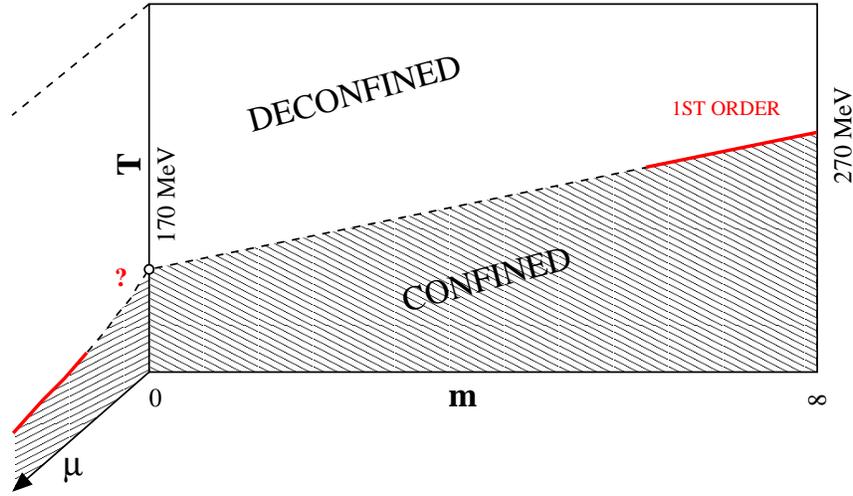}
  \caption{Picture to fixed height}
\end{figure}
At $m=0$ the chiral transition occurs, and $<\bar\psi\psi>$ is the order parameter.As $m\to \infty$ quarks decouple, the theory becomes quenched and $<L>$ is an order parameter. At generic values of $m$ neither $<L>$ nor $<\bar \psi \psi>$ are order parameters. 	The transition line corresponds to sharp variations of these quantities, or to peaks in their susceptibilities whose positions coincide  within errors.

The region below the line is conventionally called confined, the region above it deconfined.
A renormalization group analysis \cite{PW} shows that the chiral transition is first order for $N_f \geq 3$
and can be second order or first order for $N_f=2$ , depending on the fact that $U_A(1)$ symmetry is restored below or above the chiral transition.

 If the transition is second order the line at $m\neq 0$ is a crossover , implying that the deconfining transition is not order-disorder (unnatural scenario) ; if the chiral transition is instead first order 
 also the line at $m \neq 0$ is first order and the transition can be order-disorder.
  Up to now lattice results were  not conclusive on this point. New data \cite{DDP}  indicate that the transition is more compatible with first order . This is a fundamental question for our understanding of confinement and deserves further study.
  \section{Duality} 
   In lattice $QCD$  the dependence of the temperature [eq.(3)] on the coupling constant  is dictated by
   renormalization group .  $ a(\beta, m) \propto  {1\over  {\Lambda_L} } exp({\beta \over {2b_o}})$
   with $ b_o= -{1\over {4\pi^2}}({11\over 3}N-{2\over 3}N_f ) <0$  , which implies 
   \begin{equation}
   T\propto exp({N\over |b_o|}{1\over  g^2})
    \end{equation}
    Large values of $g$ (disorder) correspond to low $T$, small values of $g$ (order) correspond to high  $T$ , a peculiar behavior due to the negative sign of $b_o$ , ( asymptotic freedom), which naturally leads to  Duality.
    
    Duality is a deep concept  first  developed in statistical mechanics \cite{KW}\cite{KC}, and then extended to field\cite{SW} and string theory \cite{SD}. It applies to systems with non local, topologically non trivial 
    excitations.  
    
    Such systems admit two equivalent complementary descriptions:
    
    (1) A $\underline{direct}$ description in terms of local fields $\Phi$  whose vacuum expectation values ($vev$) 
    $<\Phi>$ are the order parameters.  The topological excitations  $\mu$ are non local. This description is convenient  in the weak-coupling regime $g < 1$.
    
     (2)  A $\underline{dual}$ description in which $\mu$ are local fields, $<\mu>$ are the (dis)order parameters, and $\Phi$ are non local . This description is convenient in the strong coupling regime $g>1$, since the dual coupling constant  $g_D$ is related to the direct one as $ g_D\approx 1/g$.
     
     Duality maps the strong coupling limit of the direct description into the weak coupling limit  of the dual one and viceversa.
     
     The prototype theory is the $2d$ Ising model \cite{KC}  which is defined on a $2d$ square lattice
     in terms of a two-valued  local field $\sigma_i= \pm 1$ with nearest-neighbor  interaction $h_{ij}= J \sigma_i\sigma_j$. The magnetization $<\sigma>$ is the order parameter  which is non zero below a Curie temperature $T_c$, and vanishes above it. The partition function depends on the parameter $\beta\equiv {J\over T}$  . The model admits topologically non trivial $1d$ excitations $\mu$, which are kinks with spin up in half space and down in the other half, or viceversa (antikinks). 
      It can be shown that also $\mu$ is a two valued variable $\mu = \pm 1$ and that the equality holds
      \begin{equation}
        K[\sigma, \beta]  = K[\mu , \beta']   
        \end{equation}
        where K is the partition function  and $sinh(2\beta)=1/sinh(2\beta') $ or  $\beta\approx 1/{\beta'}$.
        The system can be described equivalently in terms of the local variable $\sigma$ or of the dual variable $\mu$ and order -disorder are interchanged in the two descriptions. In particular for $T\ge T_c$
        the dual is ordered , $<\mu>\neq 0$ and the dual coupling is small.
        
        The idea is then to look at a dual description of $QCD$ \cite{'TH}, in terms of spacial configurations with non trivial topology $\mu$. In this description the low temperature phase which is disordered in the usual direct description should be ordered and  $<\mu>$ should be the order parameters.
        
        The problem is then to identify the dual symmetry and the corresponding topological excitations.
        If  $G$ is the gauge group topological excitations in $(3+1)d$  are obtained by breaking the symmetry to a subgroup  $H$  provided that the homotopy   $\Pi_2(G/H) $ is non trivial. Such excitations are monopoles  \cite{'TH81} . The procedure is called abelian projection\cite{'TH81}. For $SU(3)$  gauge group there are two independent monopole species.
        
        In $(2+1)d$ the relevant homotopy is $\Pi_1(SU(3)/Z_3)$ and the natural topological excitations are vortices. Vortices could be relevant to confinement also in $(3+1)d$\cite{'TH}.
        
        Understanding the relevant dual excitations and their symmetries  amounts to understand the mechanism of color confinement.
        \section{Mechanism of confinement: results and perspectives}
        The literature of the last ten years on the subject can be classified according to the guiding principles as follows.
        
        1) Search for dual symmetry.  The basic idea is that  the dual excitations  carry magnetic charge
        and condense in the confined phase producing dual superconductivity. An operator  $\mu$
        carrying magnetic charge is constructed\cite{DDPP},\cite{PR1},\cite{BARI} and its $vev$ is measured by numerical simulations 
        as a function of the temperature $T$. If the basic idea is correct  one expects  $<\mu> = 0$
        for  $T>T_c$ in the thermodynamic limit  ( infinite spatial volume) , $<\mu> \neq 0$ for $T<T_C$.
        In principle it is not possible by direct measurement to state that $<\mu>=0$ due to statistical error.
        To go around this difficulty a susceptibility  $\rho\equiv {{\partial ln(<\mu>)}\over {\partial \beta}}$
        is defined \cite{DDPP}. Since by construction $ <\mu>  _{\beta=0}=1$,
        \begin{equation}
        <\mu> = exp(\int_0^{\beta} \rho(\beta')d\beta')
        \end{equation}
        For $T>T_c$  $\rho \to -\infty$ as the volume goes large , showing that $<\mu  >$ is strictly  zero.
        For  $T<T_c$  $\rho$ becomes volume independent at large volumes and $<\mu> \neq 0$.
        At  $T\approx T_c$ ,  $\rho$ has a sharp negative peak  corresponding to a rapid decrease of
        $<\mu>$ and obeys the finite size scaling equation
        \begin{equation}
        {\rho \over { L_s^{1\over \nu}}}= \Phi(\tau  L_s^{1\over \nu})
        \end{equation}
        where  $\tau=1-{T\over T_c} $ is the reduced temperature, and $\nu$ is the critical index describing
        the divergence of the correlation length $\xi$, $\xi \propto \tau^{-\nu}$.
        Eq(9) allows to extract the value of $\nu$ , and to identify the order of the transition and the universality class.
        
        For quenched theory the location of the transition and the critical index obtained in this way are consistent with the determinations made by use of the Polyakov line \cite{PR1} \cite{PR2}.  Moreover the result is independent on the choice of the abelian projection used to define the monopoles\cite{PR3}.
        Also in full $QCD$ with $N_f=2$ the method works and the critical index $\nu$  turns out to be that of a first order transition \cite{PR4}\cite{CCD}.
        
        The net result of this approach is that the deconfining transition is an order-disorder transition: the vacuum is a dual superconductor in the confined phase, normal in the deconfined phase.
        Whatever the dual excitations they are magnetically charged in all the abelian projections.
        
        2) Approach choosing a special gauge, maximal abelian for monopoles\cite{SUZ}\cite{ITEP}, maximal central for vortices\cite {DFG} to expose non local excitations.
        
        "Surgical " analysis: eliminate somehow the non local excitations (monopoles or vortices )and check that physical properties 
        like string tension or alternative manifestations of confinement like the infrared divergence of the ghost propagator disappear. Claims follow that the mechanism of confinement is understood.
        
        3) An interesting recent development  is the exploration of the role of the center in the process of confinement\cite{BERN} to check the claim that vortices are the relevant excitations. In particular the gauge group $G_2$ which has no center  does confine color. 
        
        4) Phenomenology of monopoles in Maximal Abelian Gauge \cite{ZAK} : the idea is to guess from 
        the observation  of lattice configurations their dynamics and the possible existence of excitations 
        of different dimensionality.
        
        I think that a fair conclusion is that we do not know yet the dual excitations : we only know that
        they carry magnetic charge in all abelian projections, and that the confining vacuum is a dual      superconductor in the confined phase , a property which disappears above the deconfining transition.
        
        \section {Concluding remarks}
        A fundamental issue to understand confinement in $QCD$  is  whether the deconfining transition is order -disorder or can be a crossover , i.e. whether confinement is an absolute property of the vacuum based on symmetry or is a dynamical change of  physical quantities like $n_q$ of Eq(1) or $\sigma_q$ of Eq(2) by several orders of magnitude across a crossover. This question can possibly be settled by more extended simulations of the theory on the lattice.
        
        Investigations using an order parameter  which detects condensation of magnetic charges  show that dual superconductivity is at work as a mechanism of confinement.
        
        Direct investigation of the role of the center by comparison of different groups with the same center
        can provide useful information, as well as the study of confinement with gauge groups like $G_2$
        which have no center.
        
        We do not have yet a clear understanding of confinement.

\end{document}